# Magnetic avalanche-like behavior in disordered manganite $(Eu_{0.4}La_{0.1})(Sr_{0.4}Ca_{0.1})MnO_3$

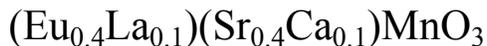

**D. S. Rana\*, and S. K. Malik[§]**

*Tata Institute of Fundamental Research, Homi Bhabha Road, Colaba, Mumbai – 400005, INDIA*

**Abstract**

The half-doped manganite, $(Eu_{0.4}La_{0.1})(Sr_{0.4}Ca_{0.1})MnO_3$, has been found to exhibit sharp step-like metamagnetic transitions below 5 K. The number of magnetic steps increases with decreasing temperature and this number suddenly rises from 3 at 2 K to ~50 at 1.7 K. The self-similar character of the multiple magnetic steps at reduced temperature identifies these steps as avalanche-like transitions. The occurrence of a multitude of magnetic steps at low temperatures, the decreasing effect of magnetic field sweep rate on the step-like metamagnetism, and the decreasing irreversibility of transition in the specific heat suggest that reduced spin-lattice coupling facilitates the transformation of antiferromagnetic to ferromagnetic state via an avalanche-like behavior.

\* Present address:      Institute of Laser Engineering,
Osaka University, Japan
Email: rana-d@ile.osaka-u.ac.jp

[§] Present address:      International Centre for Condensed Matter Physics-ICCMP,
University of Brasilia, Brasilia, Brazil
Email: skm@tifr.res.in

The half-doped $ABO_3$-type manganite systems of the general formula $R_{0.5}A_{0.5}MnO_3$ (R=trivalent rare-earth cation, A=divalent cation), exhibit large magnetoresistance and a rich magnetic phase-diagram. Such properties of these compounds depend on factors such as the bandwidth of $e_g$ electron, the A-site cation size-disorder and the $Mn^{3+}/Mn^{4+}$ ratio. A few moderately low bandwidth half-doped systems, such as $Pr_{0.5}Ca_{0.5}MnO_3$, $Nd_{0.5}Ca_{0.5}MnO_3$, $Pr_{0.5}Sr_{0.5}MnO_3$, etc., possess highly stable charge-ordered (CO) and orbital ordered (OO) states with antiferromagnetic (AFM) arrangement of spins[1-6]. Recently, the breakdown of these CO and OO states and transitions of AFM to ferromagnetic (FM) state has been shown to occur via a fascinating feature of ultra-sharp magnetic steps at low temperatures. This was first observed by Hebert et al.[7] when the Mn sub-lattice was diluted with some magnetic or non-magnetic impurities and subsequently by Fisher et al.[8] by augmenting the cation-disorder at the A-site[11]. Such low temperature metamagnetic behavior in manganite compounds is remarkable as these first order transitions display a crossover from a smooth to a sharp metamagnetic transition[7-13].

Here, we report the observation of a *multitude* of sharp metamagnetic steps in a new manganite system, namely, $(Eu_{0.4}La_{0.1})(Sr_{0.4}Ca_{0.1})MnO_3$ (ELSCMO), which possesses a large cation-disorder at the A-site. The number of magnetic steps increases with decreasing temperature and aggregates to ~50 at 1.7 K, which is a new and a novel illustration of an avalanche-like phenomenon of magnetic steps in A-site disordered manganites. The occurrence of a multitude of magnetic steps is analogous to avalanche-like systems, namely, Martensitic transitions, Barkhausen Noise and natural occurring processes such as falling of sand-dunes, occurrence of earthquakes, etc. We show that a reduced spin-lattice coupling is responsible for the occurrence of avalanche-like magnetic steps at low temperatures.

The ELCSMO sample was synthesized using the standard solid-state reaction method. Powder X-ray diffraction pattern showed that this is single-phase compound crystallizing in a distorted orthorhombic structure (space group *Pnma,* No. 62) with cell parameters: a= 5.422(1) Å, b = 7.639(1) Å and c = 5.421(1) Å. Isothermal magnetization measurements were performed using a vibrating sample magnetometer (VSM, Oxford instruments, UK). While collecting the isothermal magnetization data, the samples were warmed to 250 K each time (well above any magnetic ordering temperature) to remove the magnetic history effects. For magnetic relaxation measurements [PPMS, Quantum Design, USA], the sample was cooled to a specific temperature and a field of 80 kOe was applied for 300 seconds. After reducing this field to zero, magnetization was measured as a function of time (for 10,000 secs.) in a field of 50 Oe. The specific heat data were obtained using the relaxation method (PPMS, Quantum Design, USA).

Figures 1(a), 1(b) and 1(c) show the magnetization (M), resistance [R] and specific heat [$C_p$] vs. field [H] isotherms, respectively, at 2 K, 3 K and 5 K for $(Eu_{0.4}La_{0.1})(Sr_{0.4}Ca_{0.1})MnO_3$. The magnetic isotherms display metamagnetic transition with a crossover from a smooth broad transition at 5 K to sharp transition below it [Fig 1(a)]. For instance, at 2 K, the two steps at 45 kOe and 60 kOe are extremely sharp with a maximum width of 50 Oe. Compared to these few magnetic steps in the temperature range of 2 K - 5 K, we observe another fascinating feature of metamagnetic transitions at 1.7 K where the number of discernible metamagnetic steps increases tremendously in the same field region (Fig. 2). To show the occurrence of such multitude of steps, the related regions have been expanded and shown in Figs. 2(a) and 2(b). The number of steps at 3 K and 2 K is 2 and 3, respectively, while this number increases to ~50 at 1.7 K. This occurrence of the multitude of steps at low temperature is a unique observation of its kind in the A-site disordered manganites. The steps are also observed in resistance [R] and specific heat [$C_p$] data [Fig. 1(b) & 1(c)] at nearly the same critical magnetic field ($H_C$) as that of magnetic steps, which demonstrates the strongly correlated nature of magnetic and electronic transitions in this compound. It may also be pointed out that at 5 K, the M-H curve shows a smooth transition while the R-H and the $C_p$-H curves display step-like transitions. Such a disparity is likely to originate from the method of sweeping the field.[14] Furthermore; it is evident from the magnetization isotherms shown in Fig. 1(a) that this sample (ELCSMO) attains substantial value of spontaneous magnetization in a field of 50 kOe. Therefore, temperature dependent magnetization data in a field of around 50 kOe would be appropriate to determine the order of AFM-FM phase transition. Figure 3 shows the magnetization [M] versus temperature [T] data in fields of 50 kOe and 70 kOe, collected in cooling and warming mode. A large and a moderate hysteresis, respectively,

in fields of 50 kOe and 70 kOe, in cooling and warming data at the AFM-FM transition (around 100 K) reveal that the AFM-FM transition is of the first-order. This is consistent with similar observations in other manganites[8].

During forward field scan (0 – 9 T) of the isothermal magnetization of ELCSMO sample at 1.7 K, the large number of steps have the amplitude varying from ~0.04$\mu_B$ to ~1$\mu_B$ (Fig. 2). These steps of varying amplitudes are reminiscent of the avalanche-like character. Various avalanche-like natural processes such as earthquakes, falling of sand dunes, Barkhausen noise, Martensitic like transitions, etc. are self similar in nature and are known to obey the universal power law[15-17]. In the case of presently studied ELCSMO compound, we observe that the number of steps at 1.7 K is nearly 50. A magnified part of the 1.7 K curve shown in Figs. 2(b) & 2(c) depicts that the shape of this curve on larger and smaller scales is qualitatively similar. This illustrates the self-similar character of magnetic steps, which is a characteristic feature of the avalanche-like behavior. The observation of avalanche-like behavior in ELSCMO is an exotic feature which is seen for the first time in A-site disordered manganites. The Mn-site doped $Pr_{0.5}Ca_{0.5}Mn_{0.96}Ga_{0.04}O_3$ manganite is the only other compound, which has been reported to exhibit such multitude of steps[18]. The validity of the power law distribution of the number of steps (n) in ELSCMO [given by n($\Delta M$) $\propto$ ($\Delta M$)$^\tau$, where $\Delta M$ is the amplitude of the steps and $\tau$ is the power law exponent] cannot be ascertained because of the small value of n. This is attributed to the fact that a temperature of 1.7 K (the lowest attainable in our VSM) is not low enough to get a large statistics of steps for verification of the power law.

The physical origin for the breakdown of sharp steps into an avalanche-like multitude of steps is expected to lie in the temperature dependent weakening of spin-lattice coupling, on the basis of following experimental evidence.

i) The specific heat [$C_p$] data (Fig.1b) reveals that the irreversibility (at the transition) in the heat capacity isotherms starts decreasing with decreasing temperature and almost vanishes in 2K isotherm vis-à-vis the increasing irreversibility in magnetization with decreasing temperature (Figs. 1a & 4). According to the relation $C_p=\gamma T+\beta T^3$, the major contributions to the specific heat arise from the conduction electrons ($\gamma T$) and the lattice ($\beta T^3$). It is known that in charge-ordered insulating manganites, there is almost no electronic contribution to the specific heat which means it is the lattice (phonon) contribution which decreases with decreasing temperature. The consequent weakening of the spin-lattice coupling results in almost no irreversibility in the specific heat isotherms at 2 K.

ii) The effect of varying magnetic field sweep rate on the metamagnetic transitions in ELSCMO (Fig. 4) shows that, at 4.5 K, a smooth metamagnetic transition in 3 kOe/min sweep rate transforms to a step-like transition in 5 kOe/min sweep rate. While there is only a spread of 1 kOe in $H_C$ (of first step) at 2 K, the spread in $H_C$ at intermediate temperature of 4 K is ~3 kOe [not shown]. The effect of field sweep rate was also studied at 1.7 K but we did not observe any pronounced qualitative behavioral change[19]. These data suggest that the lower the field sweep rate, the higher is the critical field. Also, the effect of field sweep rate is more pronounced near the crossover temperature but decreases with decreasing temperature. Such a behavior is most likely associated with decreasing spin-lattice coupling at low temperatures in such a way that, at 4.5 K (for instance), in low field sweep rates, the lattice gets more time to accommodate strain. Hence, a certain amount of spin-lattice coupling facilitates the smooth metamagnetic transition. However, with the same energy, lattice does not get ample time to accommodate strain when field is swept at a larger rate of 5 kOe/min, and this results in sharp steps and multitude of steps at still lower temperatures. This indicates that the weakening of spin-lattice coupling is responsible for occurrence of multitude of steps.

iii) Magnetic relaxation, measured at temperatures of 50 K, 30 K, 10 K and 2 K, was found to decays logarithmically [Fig. 5]. The considerable relaxation observed at 30 K decreases moderately at 10 K but disappears entirely at temperatures closer to 2 K. This suggests that at low temperatures the spin moments, saturated in FM state, do not get any energy from lattice to return to their original state even after field is reduced to zero.

The above-mentioned experimental evidence suggests that at low temperatures, a reduced thermal energy and, hence, depleting spin-lattice coupling result in a crossover from broad to sharp steps and, at further lower temperatures, these few sharp steps show a breakdown into multitude of magnetic steps. This may be further understood by correlating it with strain that develops (below the crossover temperature) at the interfaces of different crystallographic structures associated with the AFM and the FM phases[20].

Due to the reduced spin-lattice coupling at low temperatures, the AFM and FM structural interfaces may get more distorted and strained and, hence, not allow the smooth growth of FM phases. However, at a certain critical magnetic field, the FM phases attain sufficient energy to overcome strain and grow in a multiple step-like catastrophic manner. A further understanding of the multiple magnetic avalanches may be envisaged in the framework of a unified picture correlating the phase coexistence to the intrinsic elastic energy and to the strain as proposed by Ahn *et al.*[21], and by invoking a correlation of strain with the reduced lattice energy, as follows. At 4.5 K, the occurrence of only one magnetic step (Fig. 4) indicates that magnetic field overcomes a long-range strain to facilitate sudden growth of a FM phase. However, owing to the decreased lattice energy (below 4.5 K) the strain between various coexisting FM and AFM phases does not retain its long-range order and, therefore, results in distinct growth of FM phases in multiple number of steps with decreasing temperature. The occurrence of multiple steps at 1.7 K suggests that the strain between various FM and AFM phases is of discrete nature causing the growth of FM phase in an avalanche-like multitude of steps.

In summary, an interesting evolution of multitude of magnetic steps at low temperatures has been observed in $(Eu_{0.4}La_{0.1})(Sr_{0.4}Ca_{0.1})MnO_3$ manganite. A large number of magnetic steps at 1.7 K and the decreasing effect of field sweep rate with decreasing temperature on sharp metamagnetic steps suggests that the reduced thermal energy and, hence, the reduced spin-lattice coupling favors the transformation of AFM to FM state via nearly an avalanche-like behavior of magnetic steps. This is a unique observation of avalanche like magnetic transitions in the A-site disordered manganite compounds and it opens up new avenues for further experimental and theoretical investigations on charge-ordered manganites.

Authors thank Prof. Mustansir Barma for useful discussions.

**Figure 1(a, b and c):**

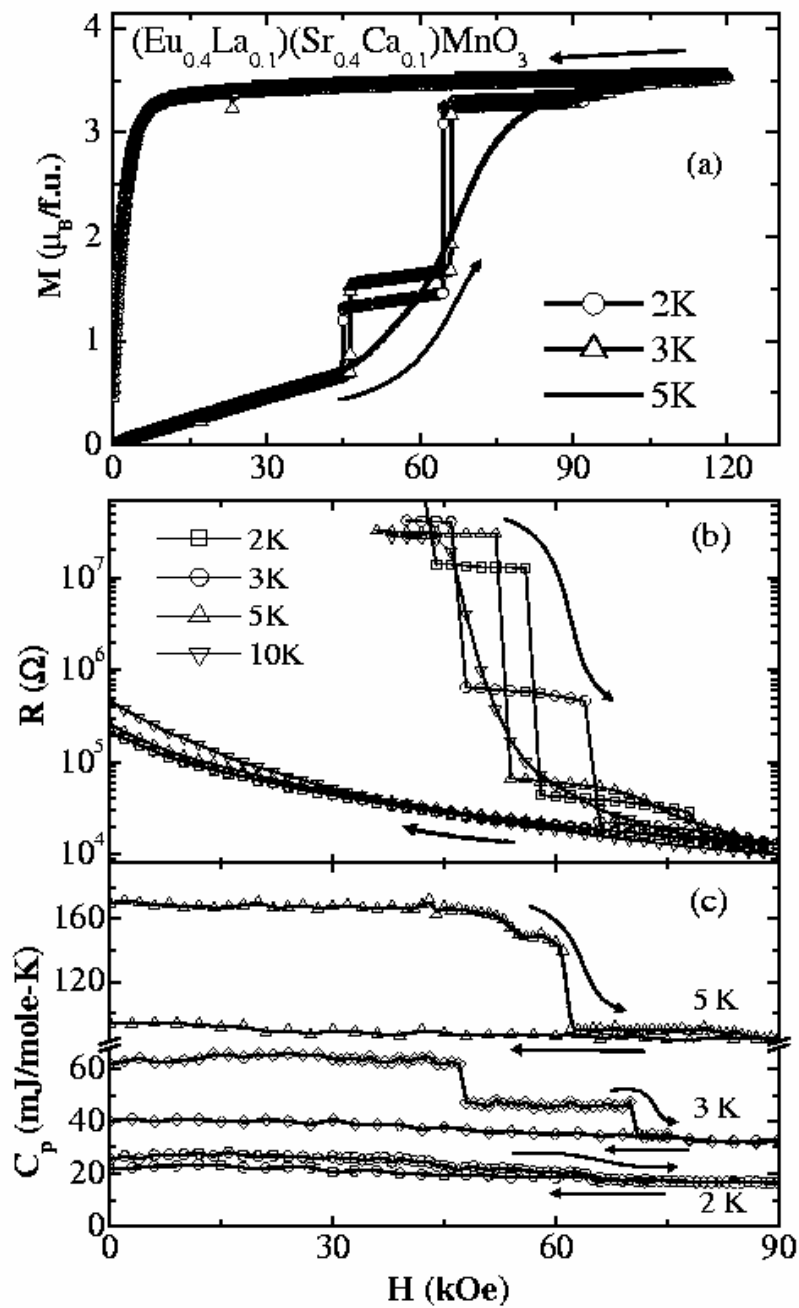

Figure 1: (a) Magnetization [M] vs. field [H] (b) Resistance [R] vs. field [H] and (c) specific heat [$C_p$] vs. field [H] isotherms at various temperatures for $(Eu_{0.4}La_{0.1})(Sr_{0.4}Ca_{0.1})MnO_3$ compound.

Figure 2:

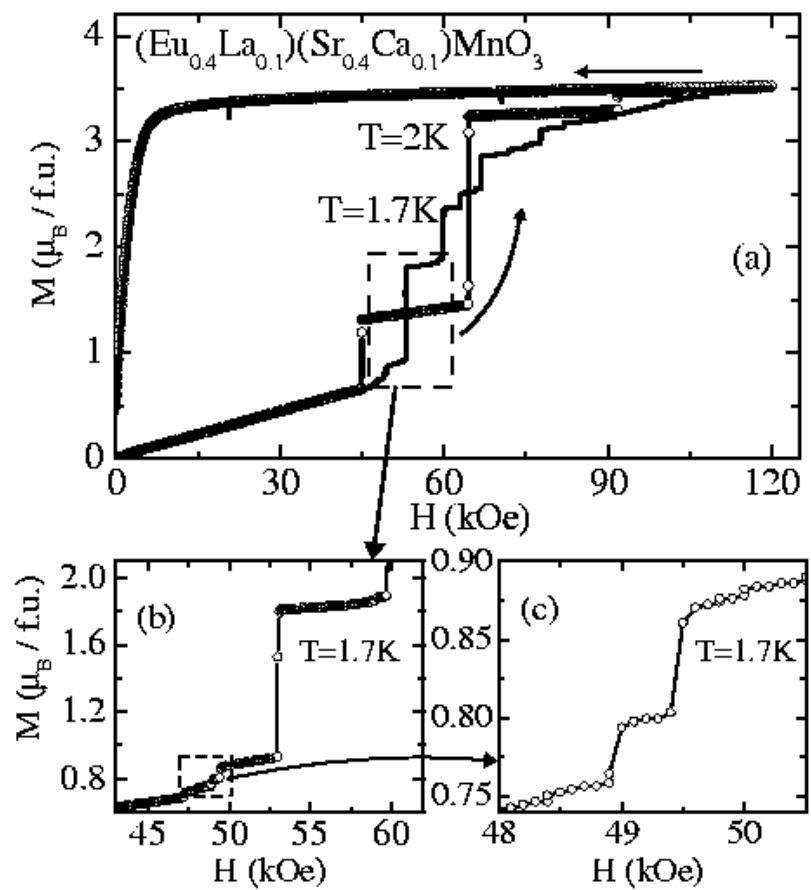

Figure 2: (a) Magnetization [M] as a function of magnetic field [H] at 2 K and 1.7 K for $(Eu_{0.4}La_{0.1})(Sr_{0.4}Ca_{0.1})MnO_3$ compound, b) a part of 1.7 K M-H curve enclosed in a box of (a) is magnified and similarly c) a part of enclosed M-H curve in a box of (b) is magnified to highlight the self-similar character of 1.7 K M-H curve.

Figure 3:

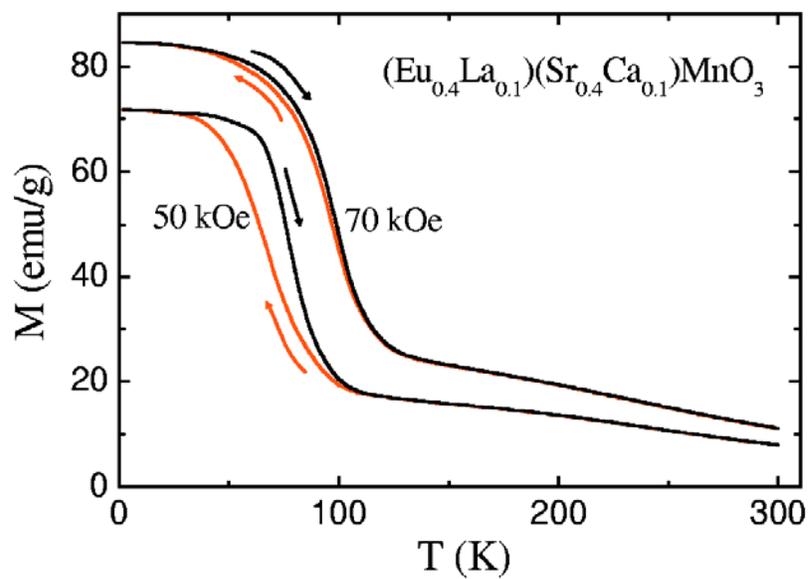

Figure 3 (color online): Magnetization [M] vs. temperature [T] in fields of 50 kOe and 70kOe for $(Eu_{0.4}La_{0.1})(Sr_{0.4}Ca_{0.1})MnO_3$ compound. The arrows indicate the cooling/warming mode for collecting the data.

Figure 4:

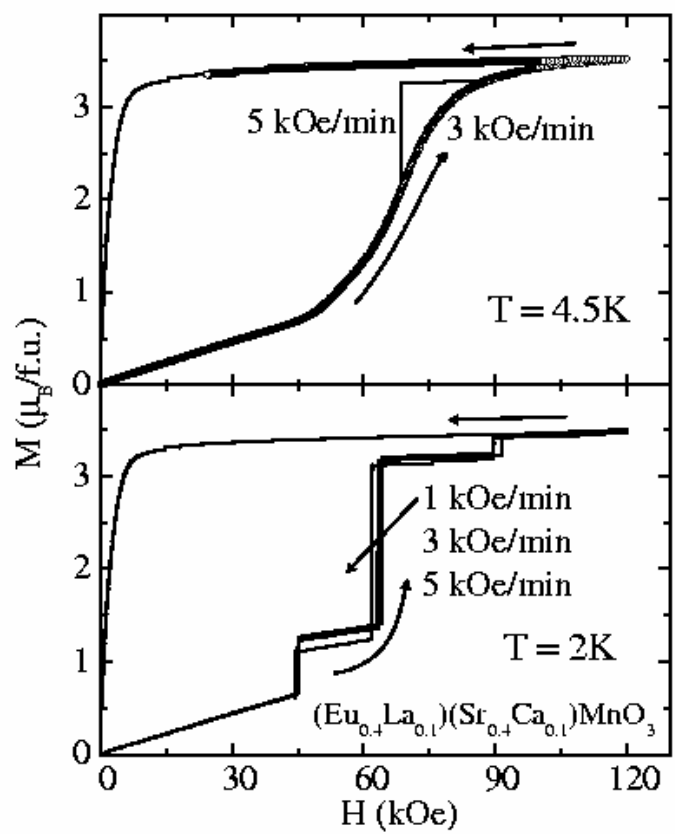

Figure 4: Magnetization [M] as a function of magnetic field [H] in varying field sweep rates of 1 kOe/sec, 3 kOe/sec and 5 kOe/sec for $(Eu_{0.4}La_{0.1})(Sr_{0.4}Ca_{0.1})MnO_3$ compound.

Figure 5:

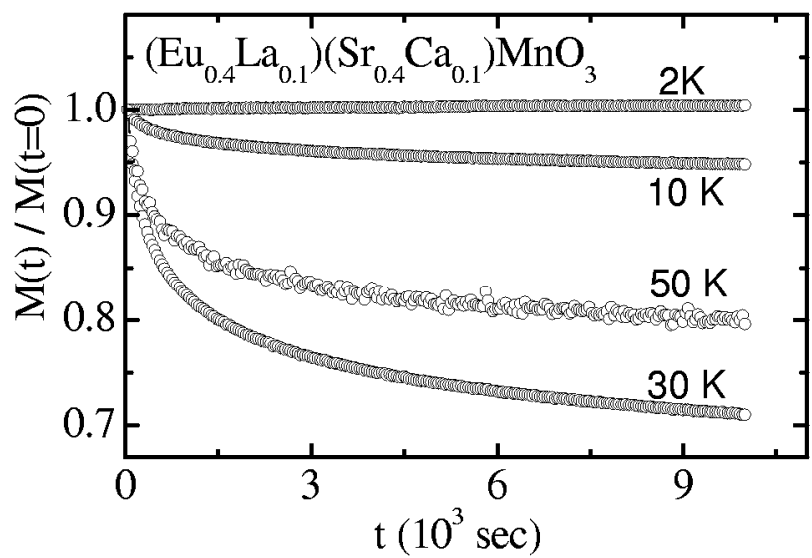

Figure 5: Normalized magnetization [M(t)/M(t=0)] versus time [t] at various temperatures.